\documentclass{cupconf}
\usepackage{graphicx}

  \checkfont{eurm10}
  \iffontfound
    \IfFileExists{upmath.sty}
      {\typeout{^^JFound AMS Euler Roman fonts on the system,
                   using the 'upmath' package.^^J}%
       \usepackage{upmath}}
      {\typeout{^^JFound AMS Euler Roman fonts on the system, but you
                   dont seem to have the}%
       \typeout{'upmath' package installed. cupconf.cls can take advantage
                 of these fonts,^^Jif you use 'upmath' package.^^J}%
      }
  \else
  \fi

  \checkfont{msam10}
  \iffontfound
    \IfFileExists{amssymb.sty}
      {\typeout{^^JFound AMS Symbol fonts on the system, using the
                'amssymb' package.^^J}%
       \usepackage{amssymb}%
       \let\le=\leqslant  
         
      }{}
  \fi

  \IfFileExists{amsbsy.sty}
    {\typeout{^^JFound the 'amsbsy' package on the system, using it.^^J}%
     \usepackage{amsbsy}}
    {\providecommand\boldsymbol[1]{\mbox{\boldmath $##1$}}}

\newsavebox{\astrutbox}
\sbox{\astrutbox}{\rule[-5pt]{0pt}{20pt}}

\newcommand\etal{\mbox{\textit{et al.}}}

\title[Lunar Radio Observatories]%
{Radio Wavelength Observatories within the Exploration Architecture}

\author[Lazio \etal]%
{J.\ns L\ls A\ls Z\ls I\ls O$^1$,
R.\ns J.\ns M\ls A\ls C\ls D\ls O\ls W\ls A\ls L\ls L${}^2$,
J.\ns B\ls U\ls R\ls N\ls S${}^3$,
L.\ns D\ls E\ls M\ls A\ls I\ls O${}^2$,
D.\ns L.\ns J\ls O\ls N\ls E\ls S${}^4$,
\and K.\ns W.\ns W\ls E\ls I\ls L\ls E\ls R${}^1$}

\affiliation{$^1$Naval Research Laboratory, 4555 Overlook
	Ave.~\hbox{SW}, Washington, DC  20375-5351 USA\\[\affilskip]
$^2$NASA, Goddard Space Flight Center, Greenbelt, MD 20771 USA\\[\affilskip]
${}^3$Center for Astrophysics \& Space Astronomy, University of Colorado, Boulder, CO 80309 USA\\[\affilskip]
${}^4$Jet Propulsion Laboratory, California Institute of Technology, Pasadena, CA 91109}

\pubyear{2007}
\volume{...}
\pagerange{1--5}
\date{2007 January~1}
\begin{document}

\maketitle

\begin{abstract}
Observations at radio wavelengths address key problems in
astrophysics, astrobiology, and lunar structure including the first
light in the Universe (the Epoch of Reionization), the presence of
magnetic fields around extrasolar planets, particle acceleration
mechanisms, and the structure of the lunar ionosphere.  Moreover,
achieving the performance needed to address these scientific questions
demands observations at wavelengths longer than those that penetrate
the Earth's ionosphere, observations in extremely ``radio quiet''
locations such as the Moon's far side, or both.  We describe a series
of lunar-based radio wavelength interferometers of increasing
capability.  The Radio Observatory for Lunar Sortie Science (ROLSS) is
an array designed to be deployed during the first lunar sorties (or
even before via robotic rovers) and addressing particle acceleration
and the lunar ionosphere.  Future arrays would be larger, more
capable, and deployed as experience is gained in working on the lunar
surface.
\end{abstract}

\firstsection 
\section{Key Science}\label{sec:lazio.science}

A lunar-based radio wavelength observatory would have two key
advantages over ground-based observatories.  First, if located on the
far side of the Moon, the observatory would be shielded from
terrestrial transmissions, both natural and human-generated.  Only a
small fraction of the radio spectrum is allocated for use by
astronomy.  Ground-based observatories are often located in remote
areas, in an effort to allow them access as much of the spectrum as
possible.  Nonetheless, many observatories are contending with an
increasing fraction of the spectrum being inaccessible.  Second, the
Earth's ionosphere is opaque at wavelengths longer than about~30~m
(frequencies below about~10~MHz).  In contrast, without a permanent
ionosphere, the surface of the Moon opens a spectral window that is
entirely inaccessible from the ground, potentially to wavelengths as
much as two orders of magnitude longer than those accessible from the
ground.  We summarize here key science that would be enabled by a
lunar-based radio observatory.

\subsection{Cosmology---Epoch of Reionization}\label{sec:lazio.eor}

In the standard hot Big Bang cosmology, the Universe started in an
initially hot, dense ionized state.  As it expanded and cooled, it
underwent a phase transition to a largely neutral state at the time of
recombination.  After recombination, baryons began to collapse into
overdense regions, eventually leading to the formation of stars and
galaxies.  Today, the collective action of stars and galaxies maintain
the Universe in a largely ionized state.  The \emph{Epoch of
Reionization} (EoR) marks this transition from neutral to ionized and
is associated with the development of structures in the Universe.  As
structures began to form and heat their surroundings, it is expected
that the excitation state of hydrogen would become decoupled from the
temperature of the cosmic microwave background.  Thus, the formation
of the first structures may be traced by the 21-cm line emission
generated as they heat their surroundings.

Observations of high-redshift quasars and the cosmic microwave
background indicate that the EoR was well underway by a redshift $z
\sim 10$ and concluded by $z \approx 6$ (\cite{bfw+01}; \cite{dcsm01};
\cite{sbo+06}).  The growth of structures, particularly during the
linear phase preceding the collapse into stars and galaxies, also
suggests redshifted hydrogen signals may be produced from structures
with redshifts as large as $z \gtrsim 50$ (\cite{lz04}; \cite{f06}).


The expected wavelength range ($\lambda > 1.5\,\mathrm{m}$ for $z >
6$, or $\nu \le 200\,\mathrm{MHz}$) is a heavily used region of the
spectrum (e.g., containing the FM radio band), with little regulatory
protection for radio astronomy.  Ground-based arrays designed to
detect the redshifted H$\,$\textsc{i} signal are being deployed, often
in the most radio-quiet regions of this planet.  However, the absence
of local transmitters may not be sufficient as ionized meteor trails
and ionospheric layers can reflect power from distant transmitters.
In contrast, the far side of the Moon is an internationally recognized
``shielded zone'' (\cite{itu05}), with emphasis given to protecting it
for observations that are ``difficult or impossible from the surface
of the Earth,'' particularly at wavelengths longer than 1~meter ($<
300$~MHz).  The far side is also protected, during a portion of the
lunar orbit, from solar radio emissions.  Terrestrial and solar
interference may ultimately be a limiting factor for ground-based
telescopes and a lunar-based telescope will be necessary to exploit
fully the hydrogen signal from the \hbox{EoR}.

\subsection{Extrasolar Planets}\label{sec:lazio.exo}

The magnetic polar regions of the Earth and the solar system giant
planets host intense electron cyclotron masers generated by solar-wind
powered electron currents.  Magnetospheric emission can aid the
understanding of extrasolar planets by providing information that will
be difficult to obtain otherwise: The existence of a magnetic field
constrains the interior of a planet while modulation of the emission
can yield its rotation rate.  For a terrestrial planet, a magnetic
field may be important for habitability, shielding the planet from the
harmful effects of energetic charged particles (e.g., \cite{ww77}).

Empirical relations for solar system planets suggest that radio
emission from extrasolar planets may be detectable over interstellar
distances (\cite{fdz99}; \cite{lfdghjh04}; \cite{s04}).
Magnetospheric emissions from the solar system planets show a rough
correspondence between planetary mass and emission wavelength, and
\emph{only Jovian radio emissions are at a short enough wavelength
that they can be detected from the ground}.  Observations above the
Earth's ionosphere will be needed to detect and study sub-Jovian mass
extrasolar planets.

\subsection{Particle Acceleration}\label{sec:lazio.particle}

Acceleration of particles to super-thermal and relativistic velocities
occurs in a variety of astrophysical environments, from the Sun and
other dwarf stars to neutron stars and black holes to quasars.  A
fundamental astrophysical problem is understanding the mechanisms and
sites of particle acceleration.  A key aspect of particle acceleration
mechanisms is the low energy population which provides the ``seeds''
from which the highest energy particles result.  These low energy
particles emit, and are best studied at, the longest wavelengths.

Within the inner heliosphere (2--10~solar radii, $R_\odot$ measured
from the Sun's center) intense electron beams are produced, at times
with particles having energies rivaling ``the energies of some
accelerated particles in the distant [quasars]'' (\cite{cb93}), and a
significant fraction of solar wind heating occurs.  Radio wave
observations and spacecraft coronagraphs, notably those on the Solar
Heliospheric Observatory (SOHO), have provided dramatic indications of
the violent, magnetically-driven activity of the Sun and its
connection to particle acceleration.  Because of its proximity and
brightness, the inner heliosphere is one of the best places to study
the fundamental physics of particle acceleration.

Solar radio bursts are one of the primary manifestations of particle
acceleration in the inner heliosphere.  Previous space-based radio
observations have been from single dipole instruments with
\emph{no imaging capabilities}.  Thus, fundamental questions remain
about particle acceleration sites.  For example, at~1~\hbox{AU},
electron acceleration generally occurs where the shock normal is
perpendicular to the magnetic field (quasi-perpendicular,
\cite{brb+99}), similar to acceleration at planetary bow shocks and
other astrophysical sites.  In the corona, the magnetic field is
largely radial, yet the existing radio observations suggest that the
acceleration site is in front of a \hbox{CME}, a
\emph{quasi-parallel} geometry.  An imaging instrument, one with even
modest angular resolution, is required to locate the sites of radio
emission, and therefore the electron acceleration.

\section{Lunar Radio Observatories}\label{sec:lazio.arrays}

Interest in placing a radio telescope on or around
the Moon pre-dates the Apollo missions (\cite{naa66}; \cite{g67}).  A
series of workshops and conferences describe scientific goals and
preliminary concepts (\cite{w86a}; \cite{bjt89}; \cite{ms90};
\cite{kw90}; \cite{swgb00}).  We begin by discussing technical aspects
of the surface of the Moon as an observatory site; the scientific
aspects are discussed above.  We then turn to how radio observatories
might be deployed, with the capability of lunar radio observatories
growing with the human presence.

\subsection{The Moon as an Astronomical Site}\label{sec:lazio.moon}

Some of the value of the Moon as an astronomical site (at long
wavelengths) is also true for a space-based array, and there have been
proposals for such arrays (\cite{w86b}; \cite{jab+00}; \cite{mgk+06}).
In contrast to free-flying telescopes at shorter wavelengths
(\cite{lym04}), the lunar surface offers two significant benefits to a
radio array.

(1)~A dipole in space responds to the full $4\pi$~sr, so a free-flying
array must image the full sky all of the time, with all of the sources
present (Sun, Jupiter, \ldots).  Limited mass budgets have restricted
proposed arrays to a small number of elements.  The resultant
challenge is to image the entire sky with an extremely sparse
aperture.  In contrast, on the Moon, the lunar surface shields
$2\pi$~sr, making the imaging problem less challenging, and imaging
algorithms developed for terrestrial interferometers can be utilized,
whereas a space array requires the development of new algorithms.
Also, it is practical to deploy a much larger number of dipoles.
(2)~In order to form an interferometer, antenna separations must be
known and maintained to a fraction of a wavelength during the
observations.  While the relevant wavelengths are large ($\sim
100$~m), station keeping does necessitate use of on-board resources.
In contrast, the lunar surface is stable with an extremely low level
of seismic activity, so antenna positions can be determined once
and then assumed constant.

\subsection{Radio Observatory for Lunar Sortie Science
	(ROLSS)}\label{sec:lazio.rolss}

\begin{figure}
\begin{center}
\includegraphics[width=0.8\textwidth]{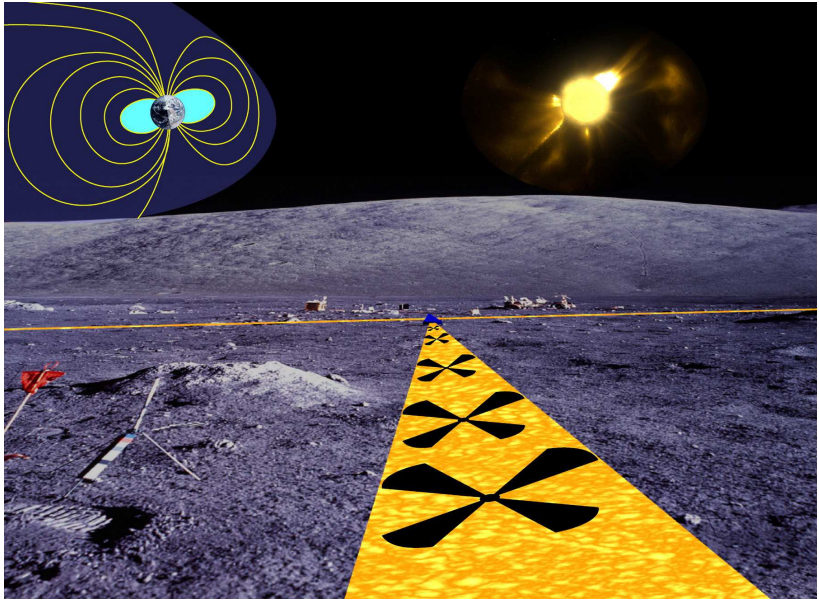}
\end{center}
\vspace*{-1ex}
\caption[]{An artist's impression of the ROLSS array deployed on the
lunar surface.}
\label{fig:lazio.rolss}
\end{figure}

The ROLSS array is a concept designed for deployment during the first
lunar sorties (Figure~\ref{fig:lazio.rolss}).  It is intended to conduct astronomical
observations of the Sun, primarily for the purpose of probing particle
acceleration mechanisms, as well as to serve as a pathfinder for
future, larger arrays.

The baseline ROLSS array consists of~3 arms arranged in a \texttt{Y}
configuration with an operational wavelength range of~30--300~m
(1--10~MHz), i.e., at wavelengths longer than can be accessed from the
ground.  Each arm hosts 16 antennas and is 500~m long, providing
approximately $2^\circ$ angular resolution at~30-m wavelength
(10~MHz).

The arms themselves consist of a polyimide film (PF) on which
electrically-short dipole antennas are deposited, and they hold the
transmission system for sending the electrical signals back to the
central processing facility, located at the intersection of the arms.
The central processing facility would select a 100~kHz sub-band within
the operational wavelength range, filter and digitize the signals,
then downlink them to the ground for final imaging and scientific
analysis.

\subsection{Future Observatories}\label{sec:lazio.future}

Observations at long wavelengths can address a number of important
science priorities, as discussed above, however, the arrays required
to conduct the requisite observations are much larger than the ROLSS
array.  A value of interferometers is that they can ``grow,'' with
scientific capability increasing as the number of antennas is
increased.  Indeed, many of the ground-based radio interferometers
have been preceded by prototypes having a much smaller number of
antennas, but which were scientifically productive themselves, and
scientific observations began with many of the ground-based radio
interferometers well before they reached their final complement of
antennas.

A strawman illustration of the staged deployment of lunar radio
interferometers is the following.
\begin{description}
\item[Stage~I (ROLSS)]~A small interferometer located on the near
side.
\item[Stage~II]~A modest-sized interferometer (e.g., 256 
dipoles spread over a few to several kilometers), possibly though not
necessarily located on the far side of the Moon.  Such an
interferometer might be capable of detecting the brightest extrasolar
planets, and verifying ground-based observations of the \hbox{EoR}.
\item[Stage~III]~A fully capable interferometer located on the far side.
\end{description}

\begin{acknowledgments}
Part of this work was carried out at the Jet Propulsion Laboratory,
California Institute of Technology, under contract with the National
Aeronautics and Space Administration.  Basic research in radio
astronomy at NRL is supported by 6.1 Base funding.

\end{acknowledgments}

\end{document}